\documentclass[
superscriptaddress,
reprint,
amsmath,amssymb,
aps,
]{revtex4-2}

\usepackage{graphicx}
\usepackage{dcolumn}
\usepackage{gensymb}
\usepackage{hyperref}
\usepackage{float}
\usepackage{outlines}
\hypersetup{colorlinks=true,pdfborder={0 0 0},}
\usepackage{enumitem}
\usepackage{upgreek}
\usepackage[normalem]{ulem}

\newcommand{\beginsupplement}{
    \setcounter{table}{0}
    \renewcommand{\thetable}{S\arabic{table}}
    \setcounter{figure}{0}
    \renewcommand{\thefigure}{S\arabic{figure}}
    \setcounter{section}{0}
    \renewcommand{\thesection}{S\arabic{section}}
    \setcounter{equation}{0}
    \renewcommand{\theequation}{S\arabic{equation}}
}

\usepackage{color,soul} %highlight to mark changes

\usepackage{xr}
\makeatletter
\newcommand*{\addFileDependency}[1]{
  \typeout{(#1)}
  \@addtofilelist{#1}
  \IfFileExists{#1}{}{\typeout{No file #1.}}
}
\makeatother

\begin{document}
\title{Efficient photon-pair generation in layer-poled lithium niobate nanophotonic waveguides}

\author{Xiaodong Shi} \thanks{These authors contributed equally.}
\affiliation{A$^\ast$STAR Quantum Innovation Centre (Q.InC), Institute of Materials Research and Engineering (IMRE), Agency for Science, Technology and Research (A$^\ast$STAR), 138634, Singapore}

\author{Sakthi Sanjeev Mohanraj}\thanks{These authors contributed equally.}
\affiliation{A$^\ast$STAR Quantum Innovation Centre (Q.InC), Institute of Materials Research and Engineering (IMRE), Agency for Science, Technology and Research (A$^\ast$STAR), 138634, Singapore}

\author{Veerendra Dhyani}
\affiliation{A$^\ast$STAR Quantum Innovation Centre (Q.InC), Institute of Materials Research and Engineering (IMRE), Agency for Science, Technology and Research (A$^\ast$STAR), 138634, Singapore}

\author{Angela Anna Baiju}
\affiliation{A$^\ast$STAR Quantum Innovation Centre (Q.InC), Institute of Materials Research and Engineering (IMRE), Agency for Science, Technology and Research (A$^\ast$STAR), 138634, Singapore}
\affiliation{Department of Physics, National University of Singapore, 117542, Singapore}

\author{Sihao Wang}
\affiliation{A$^\ast$STAR Quantum Innovation Centre (Q.InC), Institute of Materials Research and Engineering (IMRE), Agency for Science, Technology and Research (A$^\ast$STAR), 138634, Singapore}

\author{Jiapeng Sun}
\affiliation{Department of Materials Science and Engineering, National University of Singapore, Singapore 117575, Singapore}

\author{Lin Zhou}
\affiliation{Centre for Quantum Technologies, National University of Singapore, Singapore 117543, Singapore}

\author{Anna Paterova}
\affiliation{A$^\ast$STAR Quantum Innovation Centre (Q.InC), Institute of Materials Research and Engineering (IMRE), Agency for Science, Technology and Research (A$^\ast$STAR), 138634, Singapore}

\author{Victor Leong}
\affiliation{A$^\ast$STAR Quantum Innovation Centre (Q.InC), Institute of Materials Research and Engineering (IMRE), Agency for Science, Technology and Research (A$^\ast$STAR), 138634, Singapore}

\author{Di Zhu}
\email{dizhu@nus.edu.sg}
\affiliation{A$^\ast$STAR Quantum Innovation Centre (Q.InC), Institute of Materials Research and Engineering (IMRE), Agency for Science, Technology and Research (A$^\ast$STAR), 138634, Singapore}
\affiliation{Department of Materials Science and Engineering, National University of Singapore, Singapore 117575, Singapore}
\affiliation{Centre for Quantum Technologies, National University of Singapore, Singapore 117543, Singapore}

\date{May 17, 2024}

\begin{abstract}
Integrated photon-pair sources are crucial for scalable photonic quantum systems. Thin-film lithium niobate is a promising platform for on-chip photon-pair generation through spontaneous parametric down-conversion (SPDC). However, the device implementation faces practical challenges. Periodically poled lithium niobate (PPLN), despite enabling flexible quasi-phase matching, suffers from poor fabrication reliability and device repeatability, while conventional modal phase matching (MPM) methods yield limited efficiencies due to inadequate mode overlaps. Here, we introduce a layer-poled lithium niobate (LPLN) nanophotonic waveguide for efficient photon-pair generation. It leverages layer-wise polarity inversion through electrical poling to break spatial symmetry and significantly enhance nonlinear interactions for MPM, achieving a notable normalized second-harmonic generation (SHG) conversion efficiency of 4615$\%$\,W$^{-1}$cm$^{-2}$. Through a cascaded SHG and SPDC process, we demonstrate photon-pair generation with a normalized brightness of $3.1\times10^6$\,Hz\,nm$^{-1}$\,mW$^{-2}$ in a 3.3 mm long LPLN waveguide, surpassing existing on-chip sources under similar operating configurations. Crucially, our LPLN waveguides offer enhanced fabrication reliability and reduced sensitivity to geometric variations and temperature fluctuations compared to PPLN devices. We expect LPLN to become a promising solution for on-chip nonlinear wavelength conversion and non-classical light generation, with immediate applications in quantum communication, networking, and on-chip photonic quantum information processing.
\end{abstract}

\maketitle

\section{Introduction}
Correlated photon pairs are fundamental resources for photonic quantum technologies, from quantum communication and networking to sensing and computing \cite{wengerowsky2018entanglement,zheng2023multichip,xiang2011entanglement,kok2007linear,kues2017chip,couteau2023applications,flamini2018photonic}. 
They are usually generated through nonlinear optical processes such as spontaneous parametric down-conversion (SPDC) and spontaneous four-wave mixing (SpFWM).  
Integrated nanophotonic waveguides feature tight mode confinement and facilitate strong nonlinear interaction, making them well-suited for efficient photon-pair generations~\cite{wang2021integrated,moody2020chip}.
Moreover, their dense integration with various functional components in a compact chip is particularly promising for implementing scalable quantum information processors~\cite{wang2018multidimensional,wang2020integrated,pelucchi2022potential}. 

Common integrated photonic material platforms, such as silicon (Si) and silicon nitride (SiN$_x$), lack material-based second-order ($\chi^{(2)}$) nonlinearity and rely on SpFWM for photon-pair generation. 
As a $\chi^{(3)}$ nonlinear process, SpFWM usually has limited nonlinear conversion efficiency and requires long waveguides, cavities, or pulsed pumps for practical applications~\cite{shi2019multichannel,spring2013chip,xiong2015compact,xiong2011generation}. 
In contrast, SPDC is a $\chi^{(2)}$ process and can achieve higher efficiency, but demands more stringent requirements. 
Specifically, SPDC involves photons at drastically different wavelengths (e.g., visible pump light is needed for telecom photon-pair generation), making phase matching challenging, especially in nanophotonic waveguides, where geometric dispersion is severe. 
Among various $\chi^{(2)}$ materials~\cite{javid2021ultrabroadband,guo2017parametric,li2024integrated,horn2012monolithic,autebert2016integrated}, thin-film lithium niobate (TFLN) stands out as an ideal platform. It has low loss, broad transparency window, large $\chi^{(2)}$ coefficient, and most crucially, ferroelectricity that enables electrical poling~\cite{zhu2021integrated, Saravi2021_LNquantumreview}.
In TFLN, periodic poling is commonly adopted to achieve flexible quasi-phase matching (QPM) across different wavelengths. 
It has recently led to impressive results in efficient frequency conversion and non-classical light generation~\cite{lu2019periodically,javid2021ultrabroadband,hwang2023mid,park2024single}. 
However, the fabrication reliability and device repeatability of periodically poled lithium niobate (PPLN) nanophotonic waveguides remain an outstanding challenge. 
The poling quality (poling depth and duty cycle) critically depends on the fabrication conditions (e.g., poling voltage, pulse duration, temperature, electrode geometry, etc.) and directly affects the device performance. 
In addition, PPLN waveguides' phase-matching functions are sensitive to structural inhomogeneities and temperatures~\cite{zhao2023unveiling,chen2024adapted,xin2024wavelength}. 
Alternatively, modal phase matching (MPM) can achieve perfect phase matching by involving higher-order modes, where dispersions and matched wavelengths can be tailored by waveguide dimensions~\cite{wang2017second}. 
However, restricted by their symmetry properties, fundamental and higher-order modes have limited spatial overlap, resulting in low nonlinear conversion efficiency \cite{guo2016second}. 

This restriction can be resolved by breaking the spatial symmetry of the nonlinear media \cite{luo2019semi, zhang2020antisymmetric,gromovyi2023intrinsic}.
One implementation is to grow a layer of non-$\chi^{(2)}$ material (e.g., titanium oxide) on top of TFLN to form a ``semi-nonlinear'' waveguide~\cite{luo2019semi}. 
However, in this case, only half of the guided mode contributes to $\chi^{(2)}$ interaction. 
To address this problem, double-layer TFLN waveguides have been proposed and demonstrated by direct bonding of two reversely oriented $x$-cut lithium niobate (LN) films \cite{zhang2020antisymmetric, wang2021efficient,du2023highly,du2023tunable}. 
This scheme has shown a measured second-harmonic generation (SHG) efficiency of up to 5540\%\,W$^{-1}$cm$^{-2}$ and a higher theoretical SHG efficiency even exceeding that of QPM \cite{wang2021efficient}. 
Despite its high efficiency, such directly bonded double-layer wafers face several practical issues. 
Firstly, it is incompatible with integrating other functional components on the same chip, such as electro-optic modulators. 
Secondly, it was observed that wet-chemical treatments in the fabrication could introduce discontinuities on the waveguide sidewalls due to the anisotropic etching of LN waveguides with opposite crystal orientations, resulting in a high scattering loss \cite{du2023tunable}. 

In this paper, we demonstrate efficient photon-pair generation in a modal-phase-matched TFLN nanophotonic waveguide with inversely polarized layers induced by electrical poling. 
The layer-wise poling process is robust and can be locally applied to individual devices on a chip.
With the layer-poled lithium niobate (LPLN) waveguide, we measured a high normalized SHG efficiency of 4615$\pm$82\%\,W$^{-1}$cm$^{-2}$. 
Importantly, we experimentally demonstrate efficient and broadband telecom photon-pair generation in a single LPLN waveguide through a cascaded SHG and SPDC scheme. 
This scheme only requires standard telecom components (such as telecom laser and dense wavelength division multiplexer) and eliminates the need for visible pump lasers or extra SHG modules~\cite{zhang2021high,arahira2012experimental,arahira2011generation, hunault2010generation,elkus2020quantum}. 
In a 3.3 mm long LPLN waveguide, we observed broadband correlated photon pairs spanning the telecom S, C, and L bands, with a normalized brightness of 3.1$\times 10^{6}$ Hz nm$^{-1}$mW$^{-2}$, which is among the highest achieved in nanophotonic waveguides with similar configurations. 
Our device is fabrication-friendly and comparatively more stable than PPLN against variations in waveguide geometry and temperature. 
The pair-generation scheme is simple and efficient, making our LPLN photon-pair source suitable for practical applications in quantum communication and networking, as well as integrated quantum photonic information processing.

\begin{figure*}[htbp]
\centering 
\includegraphics[width = 6in]{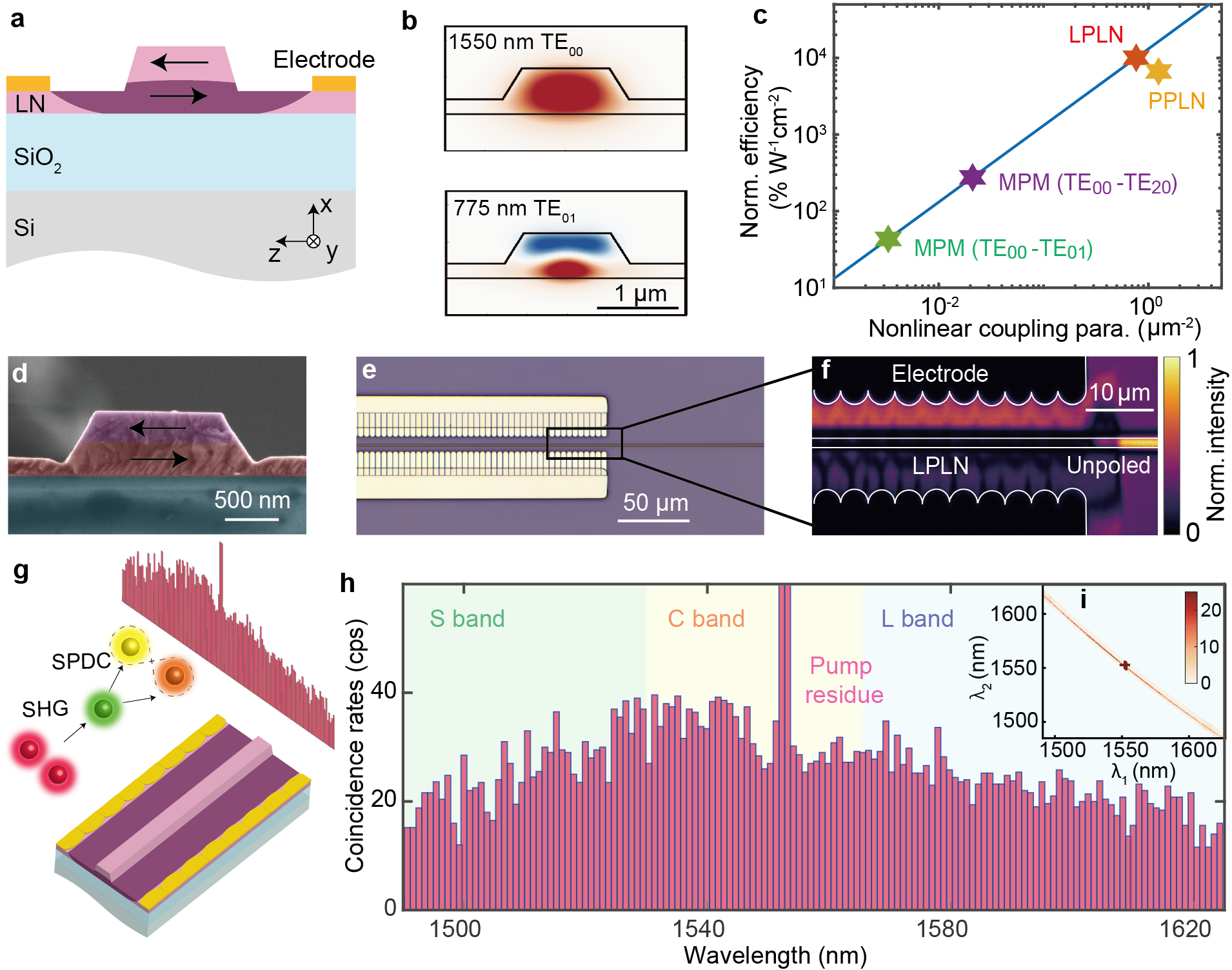}
\caption{\textbf{Layer-poled lithium niobate (LPLN) nanophotonic waveguide for efficient photon-pair generation.} \textbf{a}, Schematic of LPLN waveguide cross-section in $x$-cut TFLN. Dark and light pinks indicate inverse domain polarities. \textbf{b}, Mode profiles ($E_{z}$ component) of TE$_{00}$ mode at 1550 nm and TE$_{01}$ mode at 775 nm for modal phase matching (MPM). \textbf{c} Comparison of normalized SHG conversion efficiency among different nonlinear TFLN waveguide schemes, including LPLN (red), PPLN (yellow), MPM between TE$_{00}$ at FH and TE$_{20}$ at SH (purple), and MPM between TE$_{00}$ at FH and TE$_{01}$ at SH without poling (green). The blue line is normalized SHG efficiency versus nonlinear coupling parameter, a measure of mode overlap considering $\chi^{(2)}$ polarity distribution, with MPM condition. \textbf{d}, A false-colored scanning electron micrograph of a LPLN waveguide cross-section, showing the electrical poling induced inverse polarities. \textbf{e}, An optical micrograph of a fabricated LPLN waveguide. \textbf{f}, Confocal SHG imaging of a LPLN waveguide, where the unpoled waveguide is bright but the poled waveguide becomes dark due to the destructive interference of SH signals from the inversely polarized LN layers. \textbf{g}, Schematic of cascaded SHG-SPDC processes for photon-pair generation. \textbf{h}, Coincidence spectrum measured from 1486 nm to 1625 nm, covering telecom S, C, and L bands. \textbf{i}, Joint spectral intensity of the photon pairs. The dark cross is due to the residual pump noise.}
\label{Fig1}
\end{figure*}

\section{Results}
\paragraph{Device principle.}
Our device is designed to have MPM between the 1550 nm fundamental transverse-electric mode (TE$_{00}$) and the 775 nm first-order transverse-electric mode (TE$_{01}$). 
Due to symmetry difference (Fig.~\ref{Fig1}b), in a homogeneous waveguide, these two modes should have near-zero overlap, resulting in negligible SHG conversion efficiency (see Fig.~\ref{Fig1}c, where we use simulated SHG efficiency as a measure to evaluate nonlinear efficiencies). 
To have non-zero overlap, one needs to use an even-order mode (e.g. TE$_{20}$) at 775 nm, but the mode overlap is non-optimal. 
Here, we break the spatial symmetry of the nonlinear material by creating layer-wise inverse polarities in $x$-cut TFLN rib waveguides through electrical poling (Fig.~\ref{Fig1}a). 
This is possible as the electric field from the surface poling electrode on the LN slab is more concentrated in the lower layer due to LN's high permittivity ($\epsilon_{r,z} = 28$). 
As a result, domain inversion initiates from the bottom part of the waveguides. 
Such layer-wise polarity inversion corrects for the symmetry mismatch between TE$_{00}$ and TE$_{01}$ modes, therefore enabling a large nonlinear mode overlap. 
Since MPM is perfect phase matching, the SHG efficiency can be higher than that of QPM, which has an intrinsic $(2/\pi)^2$ penalty in SHG efficiency despite near-optimal mode overlap~\cite{boyd2020a}.  

Figure~\ref{Fig1}d shows a scanning electron micrograph of the cross-section of a fabricated LPLN waveguide. 
The false-color shading marks the layer-wise inverse polarities induced by electrical poling, extracted by intentionally immersing the waveguide in hydrofluoric acid and Standard Clean 1 (SC-1) solution (a mixture of ammonium hydroxide and hydrogen peroxide) to induce polarity-dependent etching and reveal the poled layers (see Supplementary Fig. \ref{FigS1}). 
Figure~\ref{Fig1}e shows an optical micrograph of the device, where we place electrodes with dense fingers for inducing a uniform electric field along the waveguide. 
Unlike periodic poling, we observe that the LPLN poling is relatively insensitive to poling pulse settings (voltage and number of pulses; see Supplementary Fig. \ref{FigS2}) and does not require elevated temperature or sharp electrodes (see Methods Section).
With SHG confocal microscopy, we observe that the poled section of the waveguide becomes dark (Fig.~\ref{Fig1}f). 
This is because the second-harmonic photons generated in the top and bottom layers of the LPLN waveguide are out of phase due to the reversed material polarity.
These photons destructively interfere and cancel out in far-field imaging.

To create photon pairs, we directly pump a telecom continuous-wave (cw) laser through the LPLN waveguide at its phase-matching wavelength. 
As the telecom pump traverses the waveguide, it produces SHG light, which generates SPDC simultaneously (Fig.~\ref{Fig1}g). 
Since the two processes happen in the same waveguide, the phase-matching wavelength for SHG and SPDC are automatically aligned. 
After filtering out the pump, we perform spectrally resolved coincidence counting using a pair of tunable filters and single-photon detectors. 
The photon pairs show strong frequency correlation and broad bandwidth extending the entire telecom S, C, and L bands, which is only limited by the tunable filter wavelength range (Fig.~\ref{Fig1}h, i).   

\begin{figure*}[htbp]
\center 
\includegraphics[width = 5.7 in]{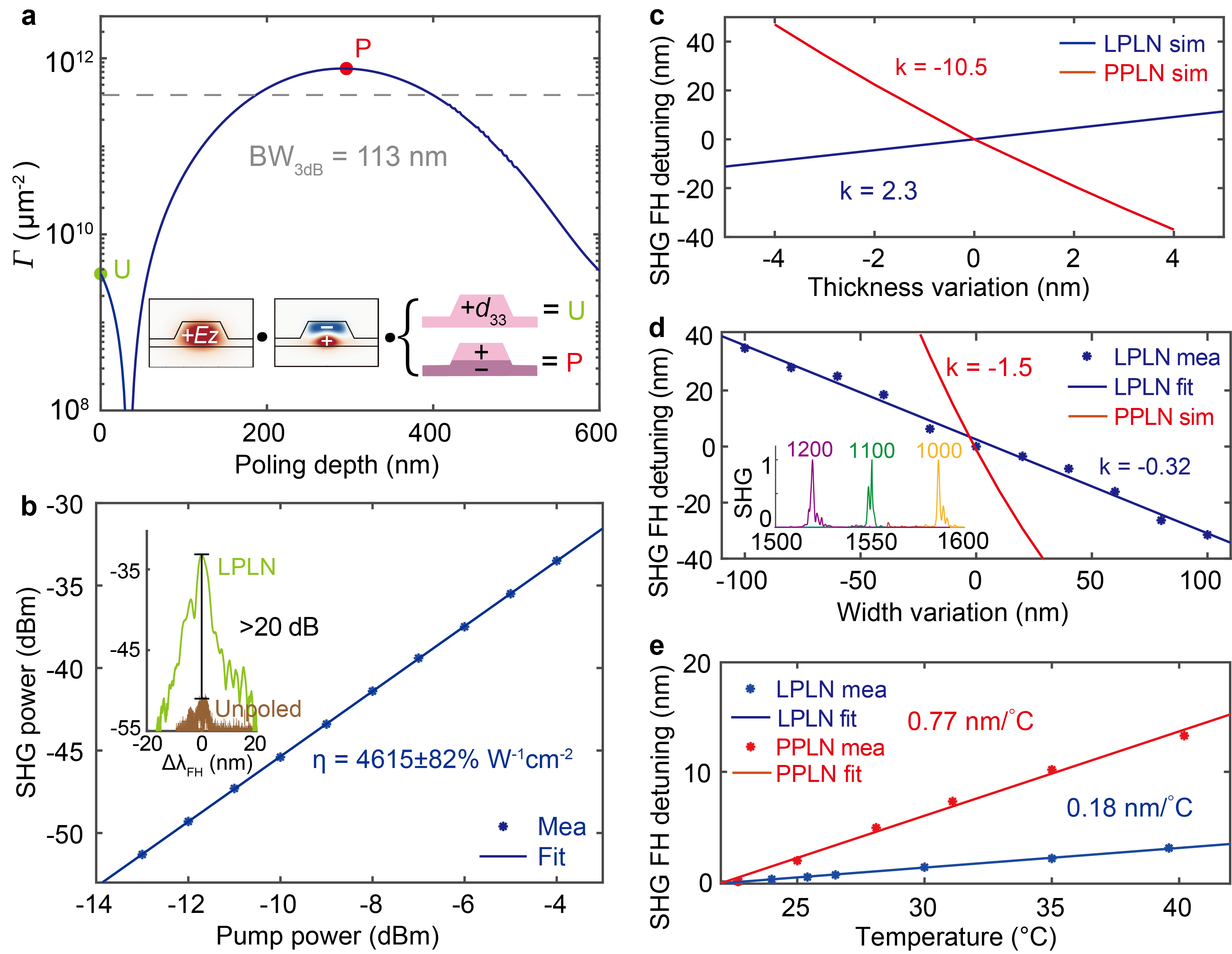}
\caption{\textbf{Numerical analysis and classical measurements of nonlinear LPLN nanophotonic waveguides.} \textbf{a}, Numerically calculated nonlinear coupling parameter ($\Gamma$) as a function of poling depth. The maximum $\Gamma$ at a poling depth of $\sim$290 nm (red dot) is more than two orders of magnitude higher than that of an unpoled LN waveguide (green dot). The inset illustrates that $\Gamma$ is essentially determined by the overlap integral of the electric fields ($E_{z}$ component) of the two involved modes and the $\chi^{(2)}$ polarity distribution in a TFLN waveguide. The gray dashed line indicates the 3 dB bandwidth (BW) of $\Gamma$, which corresponds to a poling depth range of 113 nm. \textbf{b}, Measured SHG power as a function of pump power in a 2.5 mm long LPLN waveguide. A linear fitting reveals an on-chip conversion efficiency of 4615 $\pm$ 82$\%$ W$^{-1}$cm$^{-2}$. Inset: normalized SHG spectra of a LPLN waveguide (green) and a reference, unpoled waveguide (brown), showing a $>$ 20 dB difference. Both waveguides have the same width, are fabricated on the same chip, and are tested under the same conditions. \textbf{c}, Simulated phase-matching wavelength shift (at FH) as a function of the TFLN thickness variation (nominal thickness = 600 nm) in a LPLN waveguide (blue) and a similar PPLN waveguide (red). Their thickness sensitivity is fitted to be 2.3 and -10.5 (in unit of nm wavelength shift per nm thickness change), respectively. \textbf{d}, Phase-matching wavelength shift (at FH) as a function of waveguide width variation (nominal width = 1100 nm) for LPLN (blue) and PPLN (red) waveguides, showing a sensitivity of -0.32 and -1.5, respectively. Here, the data for the LPLN waveguide is based on the measurement, and data for PPLN is from simulation. Inset:  Measured SHG spectra of LPLN waveguides with widths of 1200 nm (magenta), 1100 nm (green), and 1000 nm (yellow). \textbf{e}, Measured phase-matching wavelength shift (at FH) as a function of temperature for LPLN (blue) and PPLN (red) waveguides, showing a fitted temperature sensitivity of 0.18 nm/$^{\circ}$C and 0.77 nm/$^{\circ}$C, respectively.}
\label{Fig2}
\end{figure*}

\paragraph{Classical analysis of the nonlinear response of LPLN waveguides.}
Next, we quantitatively analyze the performances of the LPLN waveguide based on its SHG response.
The phase-matching wavelength is controlled by tailoring the waveguide dimensions. 
Here, in a 600 nm thick TFLN waveguide with 400 nm etch depth, we find that a width of 1100 nm allows the 1550 nm TE$_{00}$ mode to match 775 nm TE$_{01}$ mode (see Supplementary Fig.~\ref{FigS3}).

When the phase-matching condition is fulfilled, the normalized SHG conversion efficiency in a lossless waveguide without pump depletion can be expressed as \cite{pliska1998linear,stanton2020efficient}
\begin{equation}
   \eta = \frac{P_{\rm{SH}}}{P^2_{\rm{F H}}L^2} = \frac{8 \pi^2}{\epsilon_0 c n^2_{\rm{FH}} n_{\rm{SH}} \lambda ^2 } d^2_{\rm{eff}}\Gamma,
    \label{eq1}
\end{equation}
where $P_{\rm{SH}}$ is the generated second-harmonic (SH) power, $P_{\rm{FH}}$ is the pump power of the fundamental-harmonic (FH) wave, $L$ is the propagation length, $\epsilon_0$ is the free-space permittivity, $c$ is the speed of light, $\lambda$ is the FH wavelength, $n$ is the refractive index, and $d_{\rm{eff}}$ is the effective second-order nonlinear susceptibility ($d_{\rm{eff}} = d_{\rm{33}}$ = 27 pm/V in this work).
$\Gamma$ is the nonlinear coupling parameter between the FH and SH modes taking into account the non-uniform $\chi^{(2)}$ distribution, given by 
\begin{equation}
   \Gamma  = \frac{|\int_{{\rm LN}}{ p(x,z)\cdot (E^*_{z,{\rm FH}})^2 E_{z,{\rm SH}}{\rm d}x{\rm d}z|^2}}{{|\int_{{\rm all}}|E_{\rm {FH}}|^2 {\rm d}x{\rm d}z|^2} \int_{\rm all}{|E_{\rm SH}|^2{\rm d}x {\rm d}z}}, \label{eq2}
\end{equation}
where $p(x,z)$ denotes the $\chi^{(2)}$ polarity distribution, with $p = -1$ or 1 corresponding to polarity along $-z$ or $+z$ axis, respectively. Here, we only consider the overlap of the $z$-component of the electric fields ($E_z$) for the TE modes.

Based on numerical simulations, the nonlinear coupling parameter $\Gamma$ between the 1550 nm TE$_{00}$ mode and the 775 nm TE$_{01}$ mode is improved by two orders of magnitude in a LPLN waveguide (7.65 $\times 10^{11}$ \textmu m$^{-2}$ at an optimal poling depth of 290 nm, see Fig.~\ref{Fig2}a red dot) as compared to an unpoled one ($3.57\times10^9$ \textmu m$^{-2}$, see Fig.~\ref{Fig2}a green dot).
We also vary the poling depth and find a 3 dB bandwidth of 113 nm, which is 39\% of the optimal poling depth, suggesting that the nonlinear efficiency is reasonably tolerant against fabrication variations (Fig.~\ref{Fig2}a). 

We further calculate the normalized SHG conversion efficiency in the LPLN waveguide (at the optimal poling depth) to be  $\eta_{\rm{sim}} =$ 1.01 $\times 10^4 \%$ W$^{-1}$cm$^{-2}$, which is about twice that in a QPM-based PPLN waveguide due to $(2/\pi)^2$ penalty, and two orders of magnitude higher than other MPM-based schemes (Fig.~\ref{Fig1}c). 
Thus, LPLN holds the promise of delivering the highest $\chi^{(2)}$ conversion efficiency among other types of TFLN waveguides.

We fabricate the LPLN nanophotonic waveguides using electron-beam lithography and dry etching, followed by electrical poling to obtain the layer-wise polarity inversion in the waveguides (see Methods Section for fabrication details, and Fig.~\ref{Fig1}e and Fig.~\ref{Fig1}d for micrographs). 
Figure \ref{Fig2}b shows the measured SHG power as a function of on-chip pump power in a 2.5 mm long LPLN nanophotonic waveguide.
A linear fitting slope of 1.97 in the log-log plot confirms the quadratic relation between SH and FH powers in the undepleted-pump regime.
By calibrating out the coupling efficiencies, we extract a normalized on-chip SHG conversion efficiency of $\eta_{\rm{exp}} =$ 4615 $\pm$ 82$\%$ W$^{-1}$cm$^{-2}$.
The measured conversion efficiency is lower than the theoretical prediction, likely due to variations in waveguide dimensions and the poling depth along the waveguide, which causes fluctuations in the optimal phase-matching wavelength and the nonlinear overlap parameter, respectively.
The inset of Fig.~\ref{Fig2}b shows the measured SHG spectra of a LPLN waveguide and an unpoled reference waveguide fabricated on the same chip with the same waveguide dimensions. The LPLN waveguide shows over 20 dB higher SHG efficiency than the unpoled one, agreeing well with the theoretical predictions in Fig.~\ref{Fig2}a and Fig.~\ref{Fig1}c.

\paragraph{Comparative analysis of geometry and temperature sensitivity in LPLN and PPLN waveguides.}
A significant challenge in scaling up nanophotonic frequency converters or SPDC sources stems from the strong geometric dependence of phase-matching wavelengths~\cite{xin2024wavelength}. 
This dependency makes the fabrication of devices with identical phase-matching wavelengths difficult.
In this section, we compare the SHG phase-matching sensitivity of LPLN and PPLN waveguides. 
We first evaluate how the phase-matching wavelength changes as a function of TFLN thickness in a LPLN waveguide, and find that thicker films result in longer phase-matching wavelengths, with a simulated rate of 2.3 nm redshift per nm of thickness increase (Fig. \ref{Fig2}c).
This rate is approximately 5 times lower than that of PPLN waveguides, which has a simulated rate of 10.5 nm blueshift per nm of thickness increment.
We then measure the SHG spectra of a series of LPLN waveguides with the same length but different widths (from 1000 nm to 1200 nm in 20 nm increments) and extract their phase-matching wavelengths. 
We observe that narrower waveguides have longer phase-matching wavelengths, with a measured rate of 0.32 nm redshift per nm of width reduction  (Fig.~\ref{Fig2}d). 
The inset of Fig.~\ref{Fig2}d shows the exemplary SHG spectra with waveguide widths of 1200 nm, 1100 nm, and 1000 nm. 
The asymmetry and broadening in the measured SHG spectra are likely due to the non-uniformity in the width and height of the LPLN waveguides. 
The phase-matching sensitivity against waveguide width measured here (-0.32) is about 5 times smaller than that in the PPLN waveguide (simulated to be -1.5).
These results suggest that, compared to PPLN waveguides, LPLN waveguides exhibit reduced sensitivity to geometric variations. 

We further characterize the thermal stability of LPLN waveguides and compare it with PPLN waveguides (also fabricated on a 600 nm thick $x$-cut TFLN). 
We tune the temperature from 23~$^\circ$C to 40~$^\circ$C and measure the corresponding SHG phase-matching wavelengths in both LPLN and PPLN waveguides (Fig. \ref{Fig2}e). Both waveguides show redshifts with increasing temperature. 
The measured thermal shifting slope of the LPLN waveguide is 0.18 nm/$^{\circ}$C, about four times smaller than that of the PPLN waveguide (0.77 nm/$^{\circ}$C), suggesting better thermal stability. 

\begin{figure*}[htbp]
\center 
\includegraphics[width= 6 in]{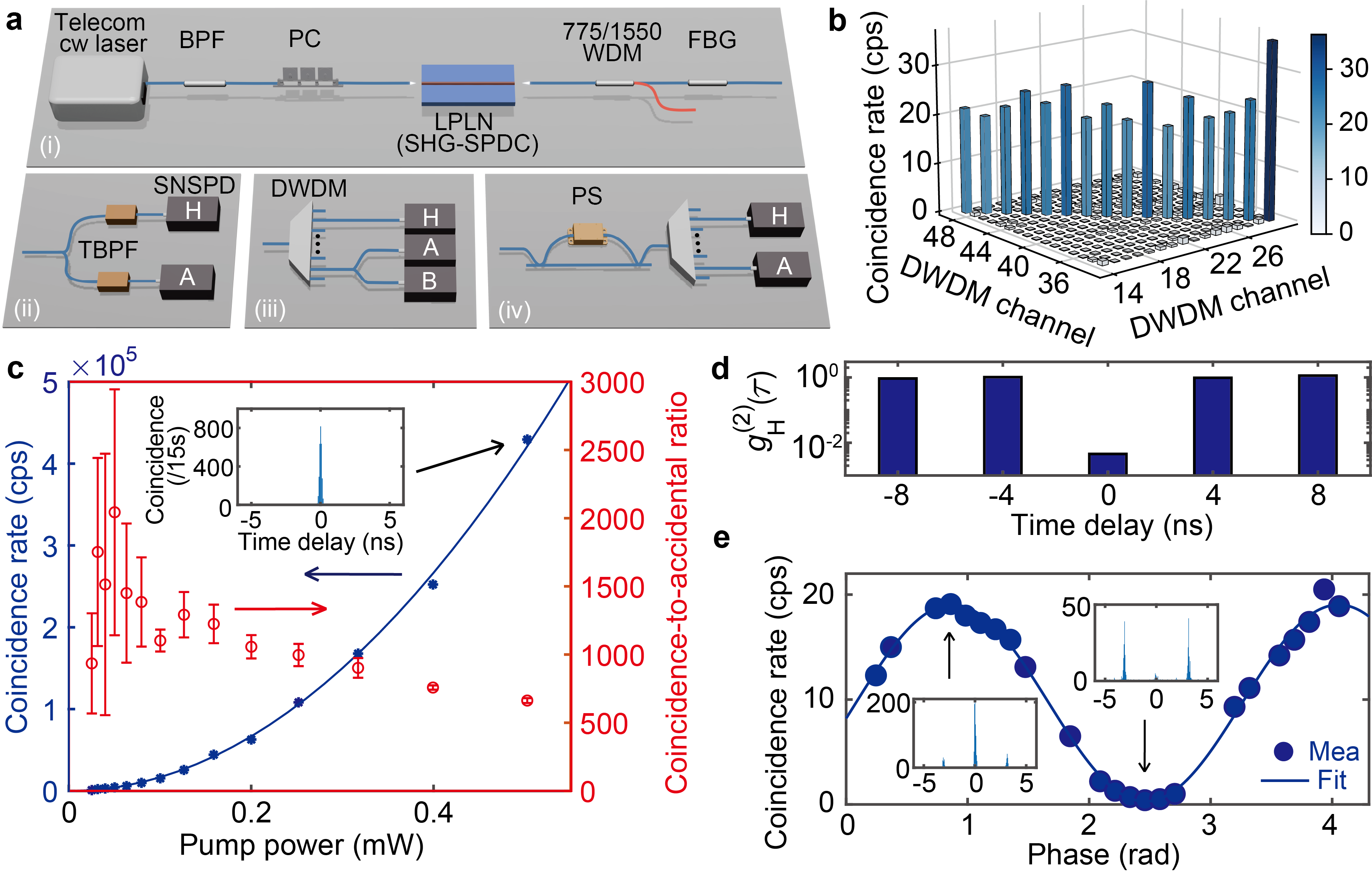}
\caption{\textbf{Non-classical characterization of photon-pair generation in a LPLN nanophotonic waveguide using a cascaded SHG-SPDC scheme.} \textbf{a}, Experimental setup for (i) photon-pair preparation, (ii) broadband photon spectral characterization, (iii) pair-generation rate, CAR, and heralded correlation measurements, and (iv) two-photon interference measurement. BPF: bandpass filter; PC: polarization controller; WDM: wavelength division multiplexer; FBG: fiber Bragg grating; TBPF: tunable BPF; SNSPD: superconducting nanowire single-photon detector; DWDM: dense wavelength division multiplexer; PS: phase shifter. \textbf{b}, Joint spectral intensity constructed by correlation measurement over 32 DWDM channels (ITU frequency Ch14-Ch29 and Ch48-Ch33), revealing strong frequency correlation through coincidences exclusively along the diagonal elements in the 16$\times$16 matrix. \textbf{c}, Measured (blue dots) and quadratically fitted (blue line) on-chip photon-pair generation rate and coincidence-to-accidental ratio (red) as a function of on-chip pump power. A raw coincidence histogram measured for 15 seconds is shown in the inset. \textbf{d}, Heralded second-order correlation function measured at various time delays at a pump power of 0.5 mW. The correlation is 0.008 $\pm$ 0.002 at zero time delay, indicating the measurements are in the single-photon regime. \textbf{e}, Two-photon interference measured at a pump power of 0.5 mW, and the measured visibility yields 98.0$\%$. Raw coincidence histograms at constructive and destructive interference are shown in the insets. Measurements in \textbf{c-e} are all done between Ch21 and Ch41.}
\label{Fig3}
\end{figure*}

\paragraph{Photon-pair generation \textit{via} cascaded SHG and SPDC.}
Finally, we use a 3.3 mm long LPLN nanophotonic waveguide for photon-pair generation through a cascaded SHG-SPDC process. 
Here, we couple a cw laser at 1552.52 nm to the chip using a lensed fiber. 
This wavelength matches the waveguide's phase-matching point and also corresponds to the International Telecommunication Union (ITU) channel, Ch31. 
The output is coupled back into a fiber and passed through a 775 nm/1550 nm wavelength division multiplexer (WDM) to filter out the SHG light and a fiber Bragg grating (FBG) for telecom pump filtering (Fig.~\ref{Fig3}a(i)). 

We characterize the photon-pair bandwidth by splitting the output into a pair of tunable band-pass filters ($\Delta \lambda_{\rm FWHM}$ = 0.6 nm) and measuring the coincidence counts as a function of wavelengths using superconducting nanowire single-photon detectors (SNSPDs, see Fig.~\ref{Fig3}a(ii)). 
Figure ~\ref{Fig1}i shows the measured joint spectrum intensity, and Fig.~\ref{Fig1}h shows the measured spectrum of the photon pairs from 1486 nm to 1625 nm, covering the telecom S, C, and L bands. 
Such broad bandwidth benefits from the low group velocity dispersion in the LPLN waveguide, resulting in a diagonally oriented phase-matching function \cite{javid2021ultrabroadband}, which is confirmed through the sum-frequency generation (SFG) measurement (see Supplementary Fig.~\ref{FigS4}). 

The broadband telecom photon-pair source is well suited for wavelength-multiplexed quantum networks. 
Here, we connect the device output to a commercial dense wavelength division multiplexer (DWDM) that matches the standard ITU frequency grid, dividing the broadband source into 32 wavelength channels. 
Figure~\ref{Fig3}b shows the measured 16$\times$16 matrix that describes the correlations among different wavelength channels. 
The correlation appears only along the diagonal elements of the matrix, demonstrating a characteristic feature of a high-dimensional quantum state and strong frequency correlation.

We pick Ch21 and Ch41, which are spectrally symmetric to the pump wavelength (Ch31), and perform detailed characterizations of the on-chip pair generation rate (PGR) and coincidence-to-accidental ratio (CAR) as a function of the pump power (Fig.~\ref{Fig3}c). 
Here, the on-chip PGR is extracted by factoring out the losses from fiber-chip coupling, filters, DWDM, and detector inefficiencies. 
It scales quadratically to pump power and reaches 0.43 Mcps at an on-chip pump power of 0.5 mW. Based on the measured channel bandwidth of $\Delta \lambda_{\rm FWHM} = 0.56$\,nm (see Supplementary Fig.~\ref{FigS5}), we estimate the normalized brightness of the photon-pair source to be $B = \frac{{\rm PGR}}{\Delta \lambda_{\rm FWHM} P^2_{\rm pump}}=3.1 \times 10^6$\,Hz\,nm$^{-1}$\,mW$^{-2}$, where $P_{\rm pump}$ is the pump power and $\Delta \lambda_{\rm{FWHM}}$ is the full-width at half-maximum (FWHM) of the signal and idler channels.
The highest CAR reaches 2043 $\pm$ 902 at 0.05 mW pump power (PGR = 4.4 kcps). 
At low pump power, the CAR is limited by dark counts; and at high pump power, it is limited by multi-photon events, Raman scattering, and residual pump photons.

SPDC sources can be used to generate heralded single photons. 
We test their purity by performing heralded second-order correlation measurement, using the detection setup in Fig.~\ref{Fig3}a(iii). 
We send the signal photons (Ch21) into a heralding detector (H) and split the idler photons (Ch41) using a 50/50 beamsplitter and measure the coincidence as a function of time delay ($\tau$) between the two detectors (A and B).  
The heralded second-order correlation is given by $g_{\rm H}^{(2)} (\tau) = \frac{N_{\rm H}N_{\rm HAB}(\tau)}{N_{\rm HA}(\tau)N_{\rm HB}(\tau)}$, where $N_{\rm H}$ is the photon counts on detector H, $N_{\rm HA/HB}$ is the coincidence counts between detector H and A/B, and $N_{\rm HAB}$ is the triple coincidence events among three detectors (H, A, and B) \cite{guo2019nonclassical}.  
Fig.~\ref{Fig3}d shows the measured $g_{\rm H}^{(2)} (\tau)$ at a pump power of 0.5 mW, where a clear anti-bunching dip with $g_{\rm H}^{(2)} (0)=0.008\pm0.002$ is observed. 
A lower $g_{\rm H}^{(2)}(0)$ is expected at lower pump power due to reduced noise photons and lower multi-photon probability. 

The cw-pumped SPDC photon pairs are naturally energy-time entangled and can be a useful resource for quantum communications. 
We perform Franson-like two-photon interference using an unbalanced Mach-Zehnder interferometer (MZI) to coherently manipulate the two-photon quantum states by applying a phase shift using a fiber stretcher (Fig.~\ref{Fig3}a(iv))~\cite{zhang2021high}. 
The zero-delay coincidence shows a sinusoidal relation to the phase shift, exhibiting a periodic transition between bunched and anti-bunched states (Fig. \ref{Fig3}e). 
The high visibility of 98.0$\%$ indicates high-quality energy-time entanglement of the photon pairs generated from the LPLN waveguide.

\begin{table*}[tbh!]
\begin{center}
\caption{\textbf{Comparison of photon-pair sources produced in a single waveguide with a telecom cw pump.} Main factors include device length ($L$), pump power ($P$), brightness, normalized brightness, as well as coincidence to accidental ratio (CAR) and heralded second-order correlation ($g_{\rm H}^{(2)}(0)$) at the corresponding brightness.}\label{tab:table1}
\begin{tabular}{  w{c}{1.3cm}  w{c}{1.1cm}  w{c}{1.0cm} w{c}{2.5cm} w{c}{3.6cm} w{c}{1.1cm} w{c}{1.1cm}}
\textbf{Platform} & {\textbf{\textit{L}}}  & {\textbf{\textit{P}}} & \textbf{Brightness}& \textbf{Normalized brightness}  & \textbf{CAR} &  {\textbf{\textit{g}}}$_{\rm H}^{(2)}$\textbf{(0)} \\
& (mm) & (mW) & (Hz nm$^{-1}$) & (Hz nm$^{-1}$ mW$^{-2}$) &  &  \\
\hline
Si \cite{clemmen2009continuous} & 11.3 & 5 & 2.0$\times$10$^6$& 7.8$\times$10$^4$ & 4 & -  \\
  Si \cite{du2024demonstration}& 8 & 11.2 & 7.1$\times$10$^6$& 5.7$\times$10$^4$ & 251 &  0.014 \\
Si \cite{guo2017high} & 10 & 1 & 1.9$\times$10$^5$& 1.9$\times$10$^5$ &  $\sim$400 &  $<$0.12 \\
 SiN \cite{choi2020correlated} & 10 & 5.0 & 4.8$\times$10$^5$& 1.9$\times$10$^4$ &  3 &  - \\
As$_2$S$_3$ \cite{xiong2011generation} & 71 & 57 & 2.5$\times$10$^6$& 7.5$\times$10$^2$ & $<$2 &  - \\
AlGaAs \cite{mahmudlu2021algaas}& 3 & 0.354 & 1.1$\times$10$^3$& 8.8$\times$10$^3$ & 21 &  - \\
This work& 3.3 & 0.5 & 7.7$\times$10$^5$& 3.1$\times$10$^6$ & 663 &  0.008 \\
\end{tabular}
\end{center}
\end{table*}

\section{Discussion and conclusion}

We benchmark our results against other reported on-chip telecom photon-pair sources produced in nanophotonic waveguides using telecom cw pump, which so far are all based on SpFWM.
The comparison in terms of brightness, CAR, and $g^{(2)}_{\rm H}(0)$ is shown in Table~\ref{tab:table1}.
We also compare the normalized brightness that accounts for the quadratic dependence of the PGR on the pump power, which is valid for both SpFWM and cascaded SHG-SPDC.
Despite its short length of 3.3 mm, our LPLN waveguide shows the highest normalized brightness while simultaneously achieving high CAR and low $g^{(2)}_{\rm H}(0)$. 
Considering that the PGR of the cascaded SHG-SPDC scheme is quartic to the waveguide length (see derivation in Supplementary Section S2) while that in SpFWM scales quadratically, we can expect drastically improved pair generation efficiency in future longer LPLN devices.

Besides $\chi^{(2)}$, LN also possesses $\chi^{(3)}$ nonlinearity, which could contribute to photon-pair generation through SpFWM.
To isolate this contribution, we shift the pump wavelength by $\sim$6 nm away from the SHG phase-matching wavelength, ensuring that the SHG-induced SPDC becomes negligible and all measured photon pairs are from SpFWM. 
We observe that the PGR from SpFWM is about two orders of magnitude lower than that from the cascaded SHG-SPDC (see Supplementary Fig.~\ref{FigS6}).
Therefore, the measured photon pairs in Fig.~\ref{Fig3} are predominantly from the cascaded SHG-SPDC process ($\sim$99$\%$).

We also experimentally compare the cascaded SHG-SPDC scheme with photon-pair generation using two separate LNPN chips, one for SHG and the other for SPDC (see Supplementary Fig.~\ref{FigS7}). 
The two-chip scheme allows more complete pump filtering, which results in a higher measured CAR approaching 4000. However, the SPDC generation efficiency is very sensitive to the optical coupling between the two chips, which is especially difficult since the SH light is in a high-order mode.
Overall, the cascaded SHG-SPDC in a single waveguide requires a simpler setup and avoids inter-chip coupling losses and mismatched phase-matching conditions.

In summary, we have developed a LPLN nanophotonic waveguide for efficient $\chi^{(2)}$ nonlinear wavelength conversion and photon-pair generation. 
The LPLN photon-pair source, which operates under a cascaded SHG-SPDC scheme, is broadband and features high normalized brightness, high CAR, and low heralded $g_{\rm H}^{(2)}(0)$, outperforming other photon-pair sources in nanophotonic waveguides using telecom cw pumps. 
Compared with traditional PPLN, the LPLN waveguide requires a simpler poling process with larger error tolerance, and its phase-matching wavelength is less sensitive to waveguide geometry and temperature variations. 
We anticipate that LPLN will be a suitable method for the future scalable production of integrated nonlinear and quantum light sources, with immediate applications in quantum communications and on-chip photonic quantum information processing.

\section*{Methods}
\paragraph{Numerical simulation.}
The waveguide effective indices and mode distributions are simulated numerically using a finite difference mode solver (Ansys Lumerical MODE). 
For the MPM design, we target phase matching between 1550 nm TE$_{00}$ and 775 nm TE$_{01}$ modes. 
We design the waveguides based on a 600 nm thick $x$-cut TFLN with an etching depth of 400 nm and an etching angle of 60$^\circ$. 
By tailoring the waveguide width, the two modes can have matched effective index (see Supplementary Fig. \ref{FigS3}). 
For PPLN waveguide simulation shown in Fig.~\ref{Fig2}c, d, we adopt the same waveguide dimension as the LPLN waveguide and keep the poling period fixed for quasi-phase matching at the corresponding zero parameter variation with phase-matching wavelength at 1550 nm, to have fair comparison with the LPLN waveguide.

\paragraph{Device fabrication.}
We fabricate the LPLN waveguides in a 600 nm thick MgO-doped $x$-cut TFLN chip.
The waveguides are patterned using electron-beam (e-beam) lithography and Ar$^+$ etching by inductively coupled plasma reactive ion etching (ICP-RIE), with hydrogen silsesquioxane (HSQ) e-beam resist as the etching mask. 
An 80 nm thick SiO$_2$ layer is deposited using ICP chemical vapor deposition (ICP-CVD) as a buffer layer for poling. 
Round-tip comb-like electrodes with a pitch of 4 \textmu m and duty cycle of 90\% are patterned using a combination of e-beam and photon lithography, followed by e-beam metal evaporation (60 nm Ni/60 nm Cr) and lift-off.
We apply a series of six 600 V, 4 ms-long electrical pulses to reverse the layer-wise polarity for LPLN waveguides, and the poling is performed at room temperature.
After poling, the SiO$_2$ buffer layer is removed using hydrofluoric acid.
A top-view optical micrograph and cross-section scanning electron micrograph of a fabricated LPLN nanophotonic waveguide are shown in Fig. \ref{Fig1}d and e, respectively.

\paragraph{Experimental details for classical characterizations.}
To extract the SHG conversion efficiency, we calibrate the fiber-to-chip coupling loss at 1550 nm using a pair of lensed fibers, and estimated the coupling loss to be 4.9 dB/facet.  
To measure the SHG power at 775 nm, we directly place a free-space-coupled visible-wavelength power meter (Si photodetector) at the chip output facet, and we assume negligible coupling loss here.
The average normalized conversion efficiency is calculated based on all the measured points shown in Fig. \ref{Fig2}b, and the uncertainty is the corresponding standard deviation.
The SHG spectra shown in the insets of Fig. \ref{Fig2}b, d are measured by synchronized sweeping of the tunable telecom pump laser while reading the Si photodetector using a data acquisition board.

\paragraph{Experimental setup for quantum characterizations.}
We use the experimental setups shown in Fig. \ref{Fig3}a to characterize the cascaded SHG-SPDC photon-pair source.
Figure \ref{Fig3}a(i) shows the setup for photon-pair generation.
A tunable telecom cw laser (Santec TLS-570) is used as the pump to stimulate the cascaded SHG-SPDC process, and its wavelength is set to 1552.52 nm, matching the SHG phase-matching wavelength and ITU Ch31.
The side-band noise from the laser is suppressed through a bandpass filter.
The pump is adjusted to be TE polarized using a polarization controller (PC) before launching into the waveguide. 
After the cascaded SHG-SPDC process in the LPLN waveguide, the pump, SHG light, and the generated photon pairs are coupled out of the chip together using a lensed fiber.
No temperature control is used during the measurements.
The SHG and pump light are filtered out through 1550 nm/775 nm WDM and FBG, respectively.
For broadband photon-pair characterization in Fig. \ref{Fig1}h, i, we use the detection setup shown in Fig. \ref{Fig3}a(ii).
The photon pairs are separated into two paths using a 50/50 beamsplitter, selected by two tunable BPFs (TBPF), and launched into two SNSPDs (ID Quantique ID281) for coincidence counting. 
During the measurement, the TBPF bandwidth is set to be 0.6 nm, and their center wavelengths sweep symmetrically to the pump wavelength.
For the narrow-band photon-pair characterization in Fig. \ref{Fig3}b-d, we use the detection setup shown in Fig. \ref{Fig3}a(iii).
Signal and idler photons are separated by DWDM with Ch21 and Ch41.
For coincidence counting, signal and idler photons are measured by two SNSPDs, H and A. 
Cascaded BPFs are used to filter the signal and idler photons to further suppress the residual pump photons.
Polarization controllers are used to optimize the photon polarization before launching into the SNSPDs in every path.
The total loss for signal and idler photons is calibrated to be 18.2 dB and 16.8 dB, respectively, by measuring fiber-chip coupling loss, optical component insertion losses, and SNSPD detection inefficiencies, in order to predict the on-chip PGR in Fig. \ref{Fig3}c.
For the heralded second-order correlation measurement, the idler photons are separated into two paths (A and B) through a 50/50 beamsplitter.
Both paths are launched into SNSPDs for photon detection, together with the signal photons (H). 
A virtual electrical time delay is added to path B in the time tagger (Swabian Time Tagger Ultra).
For the two-photon interference characterization in Fig. \ref{Fig2}e, we use the detection setup shown in Fig. \ref{Fig3}a(iv).
Before separating the signal and idler photons for coincidence counting, the photon pairs go through an unbalanced Mach-Zehnder interferometer (MZI) with a fiber-stretcher-based phase shifter (PS) in one arm. 

\section*{Acknowledgements}
This research is supported by the National Research Foundation (NRF2022-QEP2-01-P07, NRF-NRFF15-2023-0005) and A*STAR (C230917005, M23M7c0125), Singapore. The authors would like to thank Jinyi Du, Isa Ahmadalidokht, and Alexander Ling for sharing measurement equipment; Yucheng Shen and Justin Tan Zheng Jie for assistance in device measurement; Jie Deng for assistance in device fabrication.

\section*{Author contributions}
D.Z., X.S. and S.S.M conceived the idea. X.S. and J.S. designed the devices. S.S.M. and V.D. fabricated the devices. X.S., A.A.B. and S.W. performed the measurements with the help of D.Z., V.L. and A.P. L.Z. and D.Z. did theoretical analysis. D.Z., V.L. and A.P. supervised the project. All the authors discussed the results and wrote the manuscript.

\section*{Competing interests}
The authors declare no competing interests.\newline

\textit{Additional notes}---while preparing this manuscript, we noticed a conference presentation at CLEO2024 by O.~Hefti, et al. (STu3E.4), which uses a similar method to increase modal-phase-matched SHG efficiency in TFLN waveguides.

\beginsupplement
\widetext
\clearpage
\begin{center}
    \textbf{\large Supplementary Information}
\end{center}

\section{Supplementary Figures}

\begin{figure}[h]
\center 
\includegraphics[width=2.5 in]{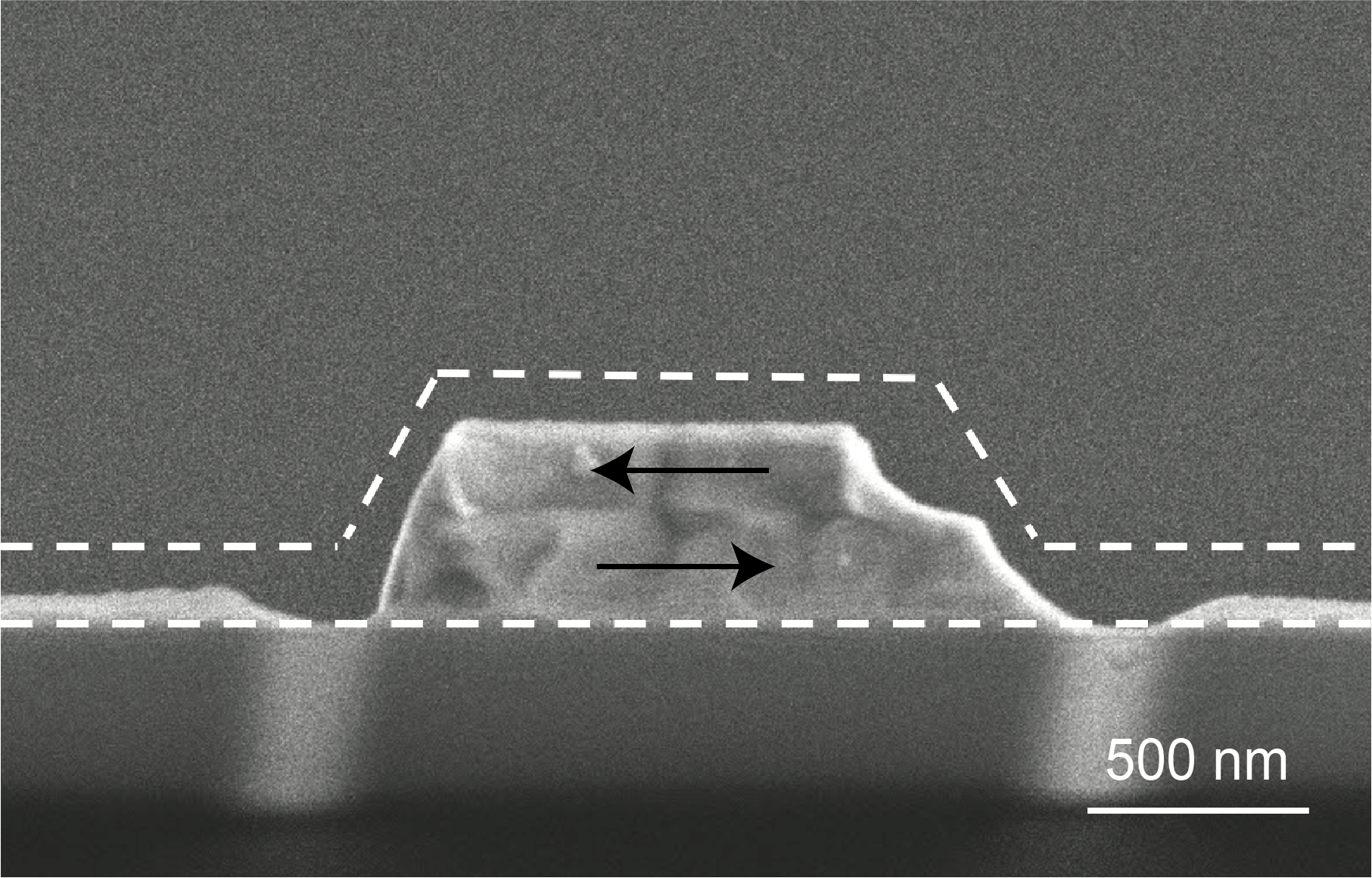}
\caption{Scanning electron micrograph of the LPLN waveguide cross section after intentional anisotropic wet etching. A clear boundary between the two domains and the discontinuity of their sidewalls can be observed, indicating layer-wise inverse polarities. The white dashed line marks the original LPLN waveguide cross section before wet etching.}
\label{FigS1}
\end{figure}

\begin{figure}[h]
\center 
\includegraphics[width=3.2 in]{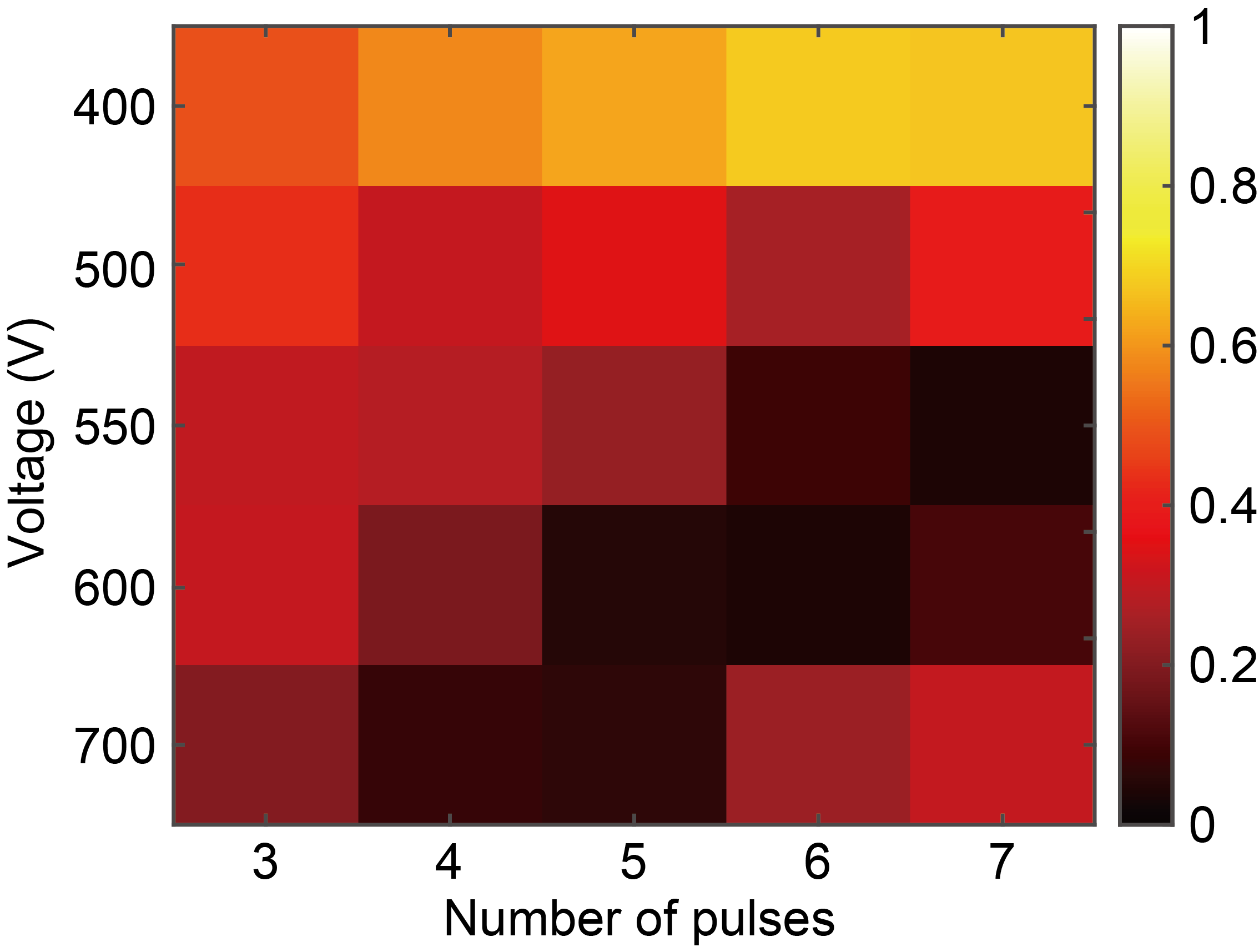}
\caption{Relation between poling depth and poling conditions (poling voltage and number of pulses). Instead of directly visualizing each device's cross-section, we use the count rates from the confocal SHG imaging (see Fig. 1f in main text)  to infer the poling depth. Here, the colorbar is the ratio between the SHG signals in the poled and unpoled waveguides. A high ratio indicates weak spatial symmetry breaking in the poled waveguides (i.e., close to being completely unpoled or fully poled), and a low ratio indicates strong layer-wise symmetry breaking (i.e., partially poled) due to the destructive interference between the two inversely polarized layers. We observe that the poling depths are relatively insensitive to the voltage and the number of pulses. The poling is performed after waveguide etching and at room temperature. We choose 6 pulses with 600 V voltage for the final device fabrication. }
\label{FigS2}
\end{figure}

\begin{figure}[h]
\center 
\includegraphics[width=3 in]{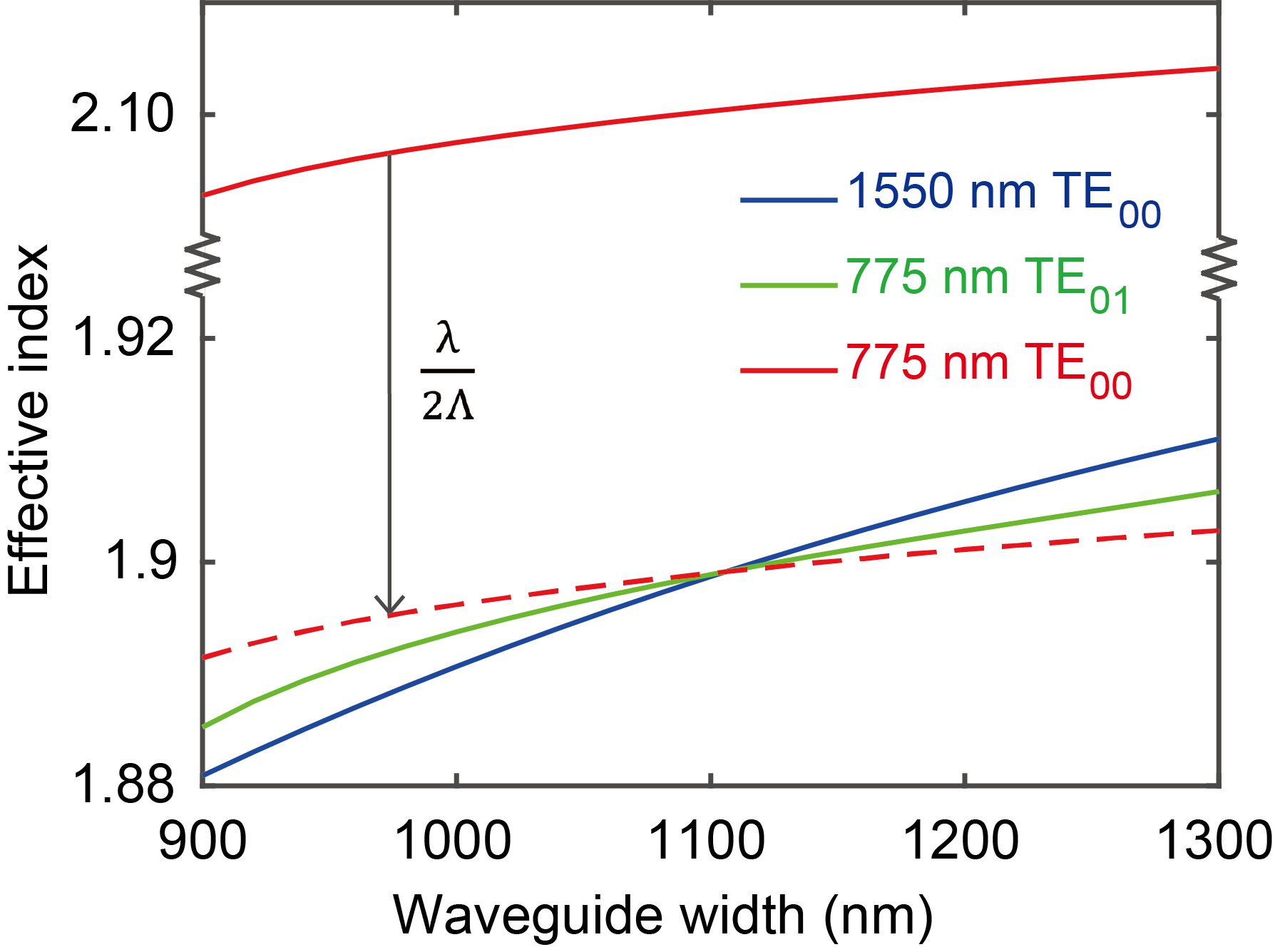}
\caption{Simulation of effective indices of the TE$_{00}$ mode at 1550 nm (blue), the TE$_{00}$ mode at 775 nm (red), and the TE$_{01}$ modes at 775 nm  (green) as a function of LN waveguide width. As the waveguide width increases, the effective index of the FH TE$_{00}$ mode exhibits a slower rate of increase compared to that of the SH TE$_{01}$ mode. The intersection point at $\sim$1100 nm between the blue and green curves signifies the phase-matching point for MPM. A larger slope difference between the FH TE$_{00}$ and SH TE$_{00}$ modes is seen compared to that between FH TE$_{00}$ and SH TE$_{01}$ modes, indicating that QPM in PPLN is more sensitive than MPM in LPLN in terms of the waveguide width variation. Here, the $\lambda/2\Lambda$ shift indicates momentum matching offered by QPM in PPLN.}
\label{FigS3}
\end{figure}

\begin{figure}[h]
\center 
\includegraphics[width=3.2 in]{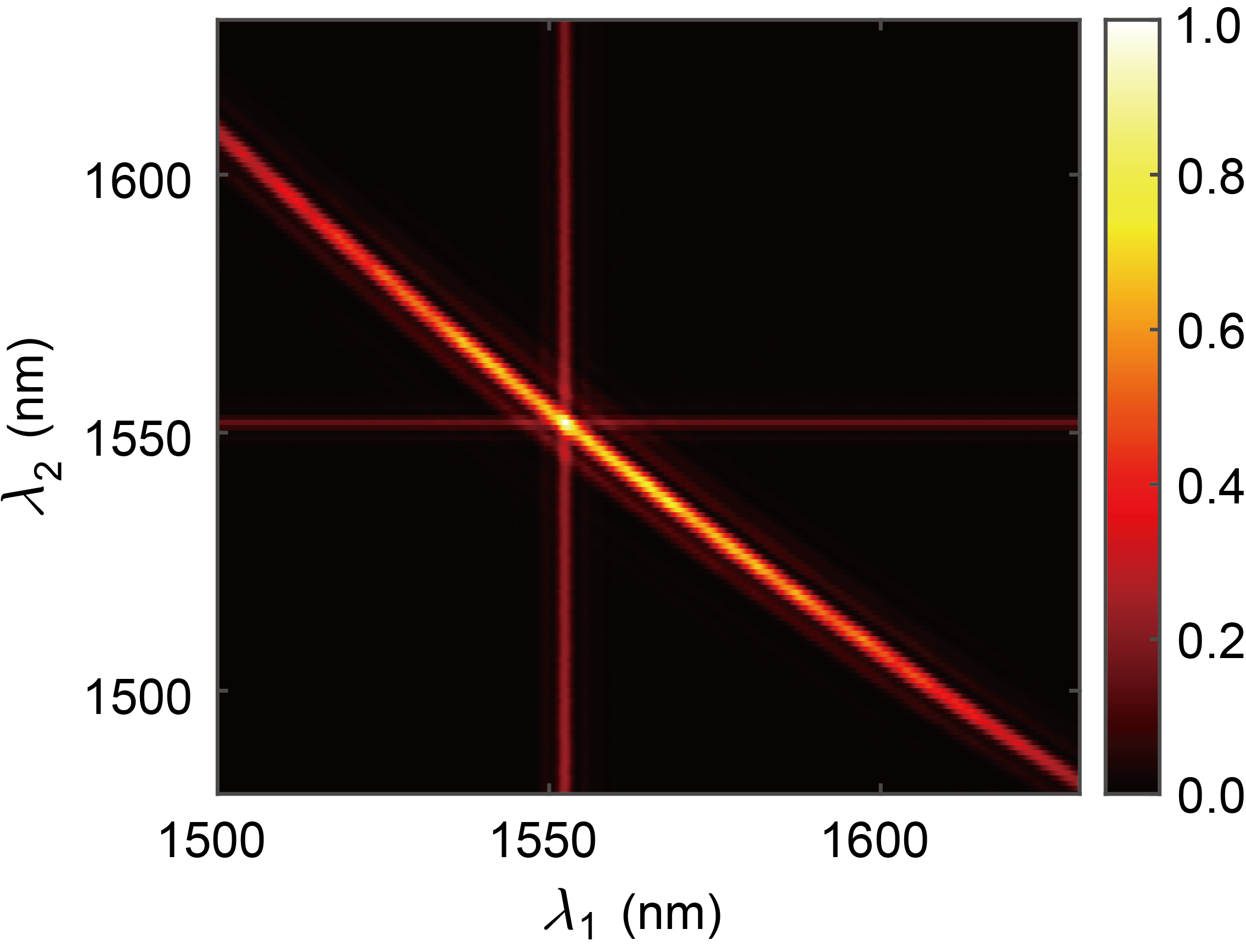}
\caption{Sum-frequency generation (SFG) phase-matching function from 1480 nm to 1630 nm, measured by sweeping two telecom cw lasers and measuring the generated SHG power. The diagonal slope follows the energy conservation line ($1/\lambda_1 + 1/\lambda_2 = 1/\lambda_{\rm pump}$), allowing broadband SPDC.}
\label{FigS4}
\end{figure}

\begin{figure}[h]
\center 
\includegraphics[width=3 in]{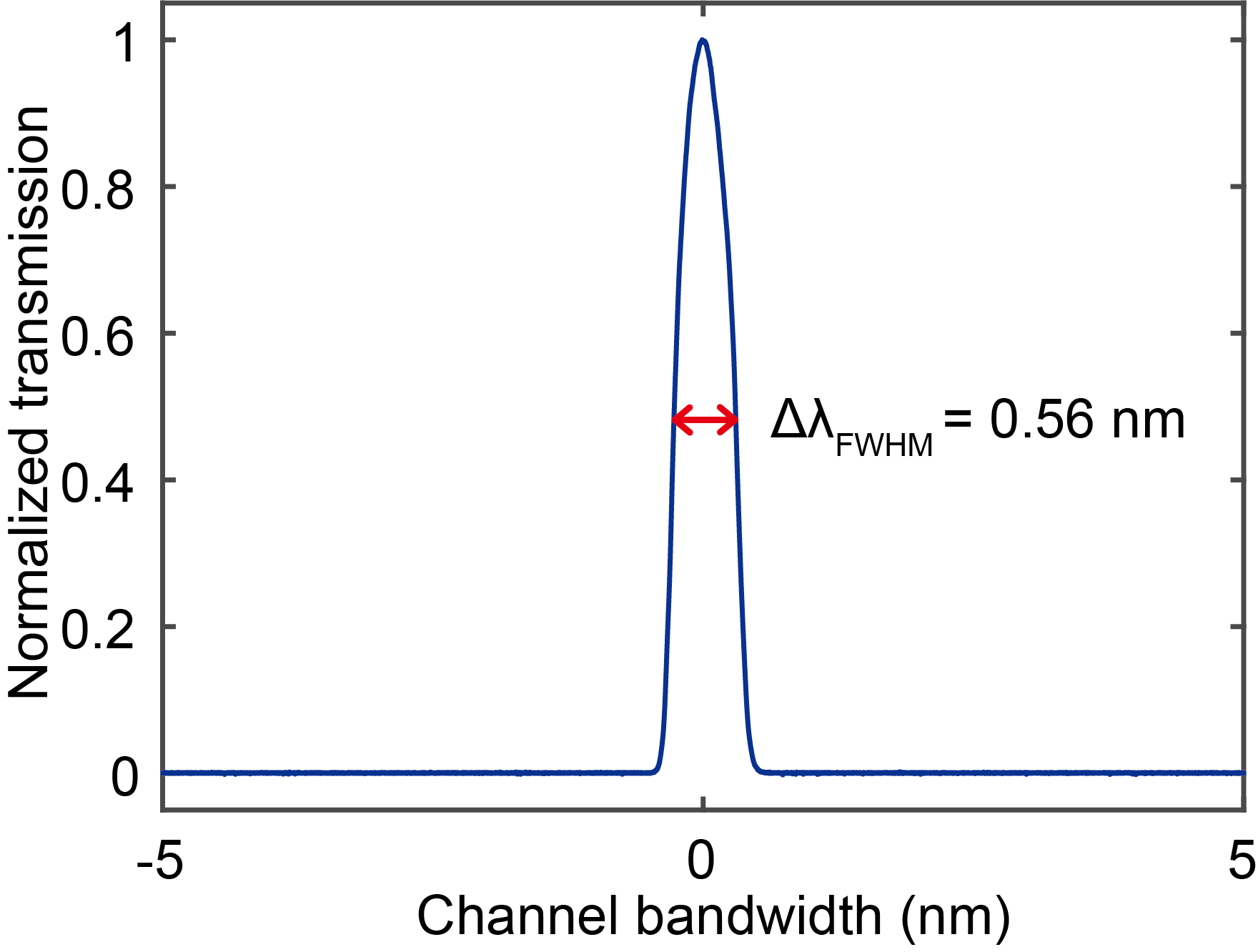}
\caption{Measured transmission spectrum of the DWDM channel used in the photon-pair measurement, showing a FWHM bandwidth of 0.56 nm. This bandwidth is used to calculate the SPDC brightness.}
\label{FigS5}
\end{figure}

\begin{figure}[h]
\center 
\includegraphics[width=3 in]{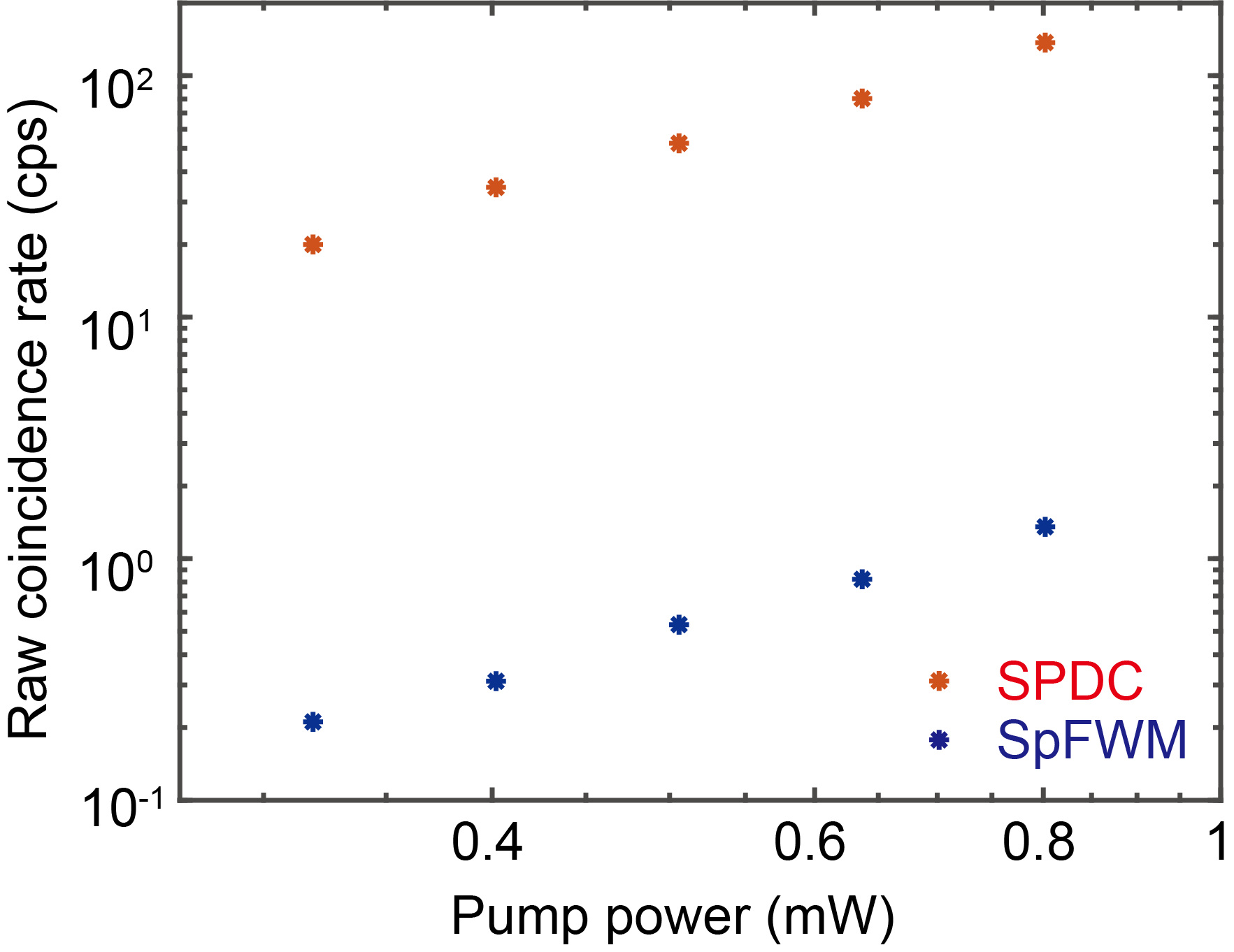}
\caption{Measured off-chip, raw photon-pair generation rates from SPDC (red) and SpFWM (blue) as a function of on-chip pump power. The PGR from SPDC is about two orders of magnitude higher than that from SpFWM.}
\label{FigS6}
\end{figure}

\begin{figure*}[h]
\center 
\includegraphics[width = 6.2 in]{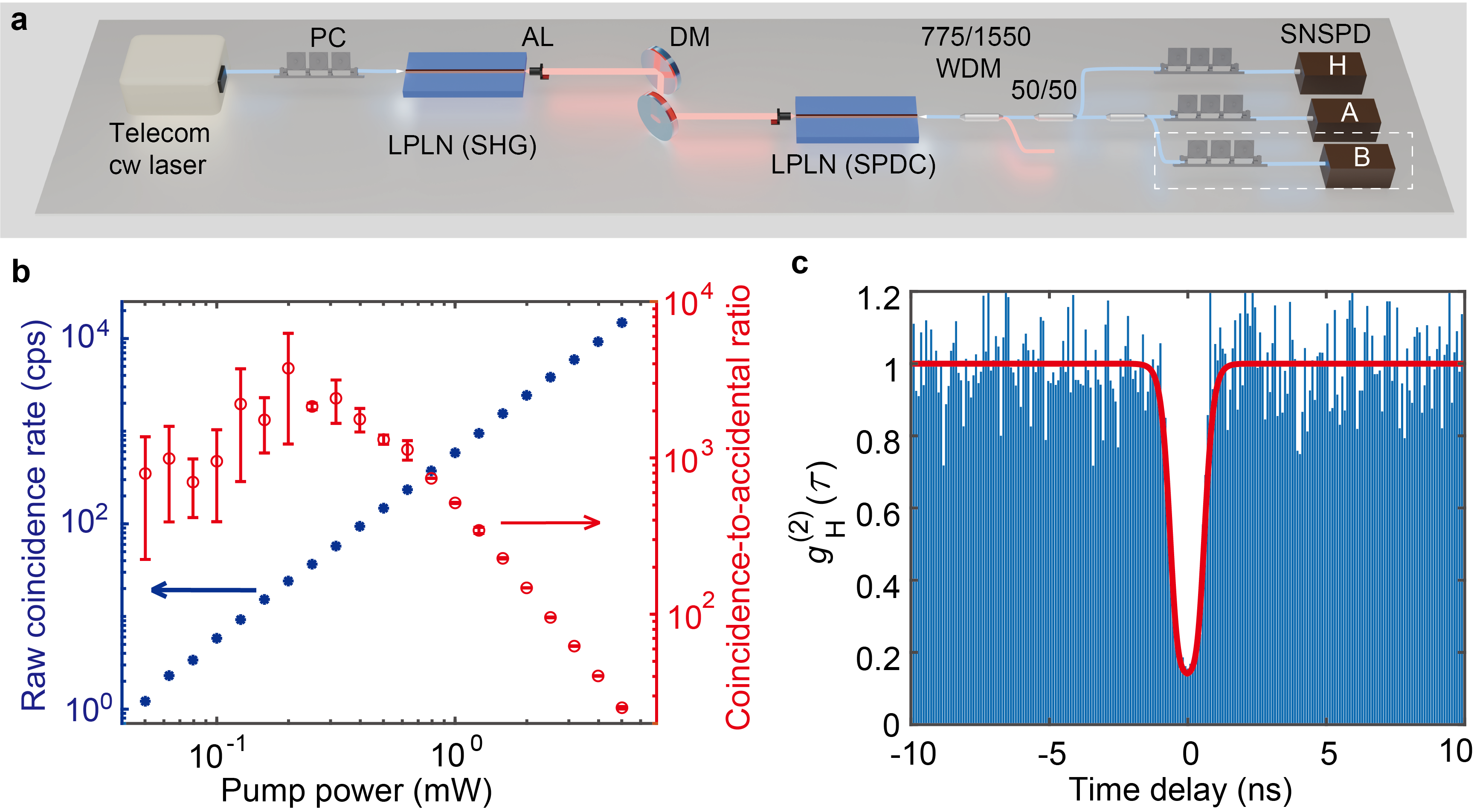}
\caption{Photon-pair generation with two LPLN chips, one for SHG and the other for SPDC. \textbf{a}, Experimental setup. The SH light, generated in the first chip, is coupled out from the first chip and then into the second chip through a pair of aspheric lenses (AL). The telecom pump light is filtered in between using dichroic mirrors (DM). SPDC photon pairs generated in the second chip is coupled out using a lensed fiber, and the FH light is filtered out through a 1550 nm/775 nm WDM. Signal and idler photons are separated using a 50/50 beamsplitter and sent to SNSPDs for coincidence counting with H and A, and heralded second-order correlation function measurement with H, A, and B.
\textbf{b}, Off-chip, raw measured photon-pair generation rate (blue) and CAR (red) versus on-chip pump power. \textbf{c} Measured (blue) and fitted (red) heralded second-order correlation function versus time delay at a pump power of 5.0 mW, and it is measured to be 0.014 at zero time delay, indicating the measurements are operated in the single-photon regime.}
\label{FigS7}
\end{figure*}

\clearpage

\section{Scaling laws for cascaded SHG and SPDC process}

In this section, we use a simplified classical model to derive the scaling laws for the cascaded SHG-SPDC photon-pair generation process. We note that this model does not capture the quantum nature of SPDC and is not sufficient to predict the absolute pair generation rate or dynamics. In a cascaded SHG and non-degenerate SPDC process based on MPM without phase mismatch, the coupled amplitude differential equation can be expressed as
\begin{align}
\frac{\partial A_1}{\partial z} &= i\frac{\omega^2_1}{c^2k_1}d_{\rm eff}A^*_1(z)A_2(z)\label{eqS1}, \\ 
\frac{\partial A_2}{\partial z} & = i\frac{\omega^2_2}{c^2k_2}d_{\rm eff}A_1(z)A_1(z)+i\frac{\omega^2_2}{c^2k_2}d_{\rm eff}A_{\rm{s}}(z)A_{\rm{i}}(z)\label{eqS2},\\ 
\frac{\partial A_{\rm{s}}}{\partial z} & = i\frac{\omega^2_{\rm{s}}}{c^2k_{\rm{s}}}d_{\rm eff}A^*_{\rm{i}}(z)A_2(z)\label{eqS3},\\ 
\frac{\partial A_{\rm{i}}}{\partial z} & = i\frac{\omega^2_{\rm{i}}}{c^2k_{\rm{i}}}d_{\rm eff}A^*_{\rm{s}}(z)A_2(z)\label{eqS4}, 
\end{align}
where $\omega$ and $k$ are frequency and wavevector, respectively, and the indices 1, 2, s, and i indicate the parameters at FH, SH, signal and idler frequencies, respectively. Following energy conservation, we have $\omega_2 = 2\omega_1$ and $\omega_2 = \omega_{\rm{i}} + \omega_{\rm{s}}$.
Furthermore, we assume the FH pump is in a non-depletion regime, and the FH and SH fields are much stronger than the signal and idler fields.
Hence, $A_1$ can be treated as a constant, and Eq.~\ref{eqS2} is simplified to
\begin{equation}
\frac{\partial A_2}{\partial z} = i\frac{\omega^2_2}{c^2k_2}d_{\rm eff}A_1A_1.
\label{eqS5}
\end{equation}
Combining Equations \ref{eqS3}-\ref{eqS5}, we have
\begin{equation}
\frac{\partial^2 A_{\rm{s}}}{\partial z^2} - \frac{1}{z}\frac{\partial A_{\rm{s}}}{\partial z}-\frac{z^2}{g^4}A_{\rm{s}} = 0,
\label{eqS6}
\end{equation}
where $g = \frac{(k_1^2 k_{\rm{i}} k_{\rm{s}})^{1/4} c^2}{(\omega_{\rm{s}} \omega_{\rm{i}})^{1/2} \omega_2 d_{\rm eff} A_1}$.
%$g = (\frac{\omega_2^4\omega_s^2\omega_i^2}{k_1^2k_ik_sc^8}d_{\rm eff}^4A_1^4)^{-\frac{1}{4}}$
%A_s(0) =0 \,\,,A'_s(0) = 0, 
Eq.~\ref{eqS6} has a general solution in the form of  
\begin{equation}
A_{\rm{s}}(z) \propto C_1\sinh(z^2/2g^2) + C_2\cosh(z^2/2g^2) \propto z^2/2g^2+\mathcal{O}(z^4),
\end{equation}
where $C_1$ and $C_2$ are constants and the last approximation assumes $z \ll g$. For a short waveguide of length $L$, we can expect the SPDC rate in the cascaded SHG-SPDC process to scale as $P_{\rm{s}} \propto P_1^2 L^4$, where $P_{\rm{s}} = |A_{\rm{s}}|^2$ and $P_1 = |A_1|^2$. The quadratic relation between the pair generation rate and the pump power agrees well with the measurement in Fig. \ref{Fig3}c.
The quartic relation between the signal power and the propagation length indicates that increasing the waveguide length is an effective means to further increase the pair generation efficiency. As a comparison, the photon-pair rate in standard SPDC with SH pump scales as $\propto P_{\rm pump} L^2$, and that in SpFWM with telecom pump scales as $\propto P_{\rm pump}^2 L^2$.

\end{document}